\documentclass[preprint,12pt]{elsarticle}
\usepackage[utf8]{inputenc}
\usepackage{cmap}
\usepackage{amssymb}
\usepackage{amsmath}
\usepackage{amsthm}

\usepackage{graphicx}
\usepackage{doi}
\usepackage[linesnumbered,ruled]{algorithm2e}
\usepackage{empheq}
\usepackage{multirow}
\usepackage{lineno}
\usepackage{scalerel}

\journal{Pattern Recognition}

\begin{document}

\begin{frontmatter}


\title{
Subword Dictionary Learning and Segmentation Techniques for Automatic Speech Recognition in Tamil and Kannada}


\author{Madhavaraj A\corref{cor1}\fnref{label1}}
\ead{madhavaraja@iisc.ac.in}
\author{Bharathi Pilar\fnref{label2}}
\ead{bharathi.pilar@gmail.com}
\author{Ramakrishnan A. G.\corref{cor1}\fnref{label1}}
\ead{agr@iisc.ac.in}
\fntext[label1]{Electrical Engineering, Indian Institute of Science\\
Bangalore, Karnataka, India}
\fntext[label2]{University College Mangalore, Karnartaka, India}
\cortext[cor1]{Corresponding Authors}

\begin{abstract}
We present automatic speech recognition (ASR) systems for Tamil and Kannada based on subword modeling to effectively handle unlimited vocabulary due to the highly agglutinative nature of the languages. We explore byte pair encoding (BPE), and proposed a variant of this algorithm named extended-BPE, and \textit{Morfessor} tool to segment each word as subwords. We have effectively incorporated maximum likelihood (ML) and Viterbi estimation techniques with weighted finite state transducers (WFST) framework in these algorithms to learn the subword dictionary from a large text corpus. Using the learnt subword dictionary, the words in training data transcriptions are segmented to subwords and we train deep neural network ASR systems which recognize subword sequence for any given test speech utterance. The output subword sequence is then post-processed using deterministic rules to get the final word sequence such that the actual number of words that can be recognized is much larger. For Tamil ASR, We use 152 hours of data for training and 65 hours for testing, whereas for Kannada ASR, we use 275 hours for training and 72 hours for testing. Upon experimenting with different combination of segmentation and estimation techniques, we find that the word error rate (WER) reduces drastically when compared to the baseline word-level ASR, achieving a maximum absolute WER reduction of 6.24\% and 6.63\% for Tamil and Kannada respectively.

\end{abstract}


\end{frontmatter}


\section{Introduction}

Research on ASR has brought lot of innovations over the last two decades. Handling unlimited vocabulary is one of the important research areas in this field. Traditional ASR models use words as lexical units which is not suitable for highly agglutinative and inflective languages ~\cite{bhref11}. Hence, we employ subword modeling approach to segment each word into subword units and use them as lexical units for training our ASR models, This subword-based ASR fares better than the word-based ASR approach in terms of reducing the WER and OOV rates, and the model complexity due to small subword vocabulary size. It also allows us to build n-gram subword language models to cover millions of words ~\cite{bh_ref10}. Various subword modeling approaches are available in the literature for European languages which uses n-grams and morphological analyzers for subword dictionary learning. Morphological analyzer like \textit{Morfessor} has been used in ~\cite{bhref13},~\cite{bhref14}, whereas syllables and rule-based algorithms have been used in ~\cite{bhref15},~\cite{laureys2002hybrid} for building subword ASRs. Data-driven algorithms that uses minimum description length and maximum likelihood estimation are used to build subword-ASRs in ~\cite{hacioglu2003lexicon} and ~\cite{arisoy2007language}. In addition to directly applying morpheme-based n-grams, the morphological and lexical information are combined and applied as factored and joint lexical–morphological language model in \cite{kirchhoff2006morphology} and ~\cite{sarikaya2007joint} respectively.

In this paper, we present novel word segmentation and subword dictionary learning algorithms to build subword-ASRs and report its performance for Tamil and Kannada speech recognition tasks. The rest of the paper is organised as follows. Section \ref{secDataBaseline} describes the baseline ASR system, dataset and tools used in our experiments. In section ~\ref{sec:subDict}, we describe the algorithmic and mathematical tools like byte pair encoding and morphological analyzer that have been used to construct the subword dictionary from the text corpus. Statistical formulation of the word segmentation techniques with maximum likelihood and Viterbi criteria are elucidated in sections ~\ref{sec:statml} and ~\ref{sec:statvit} respectively. Sections ~\ref{sec:ml} and ~\ref{sec:Viterbi} respectively describe the implementation of ML and Viterbi segmentation methods using weighted finite state transducers (WFST) framework. In section \ref{sec:subASR}, we describe the entire subword ASR system training, testing and post-processing pipelines. Finally, we present the experimental results in section \ref{sec:exp} and conclude in section \ref{sec:conc}.

\section{Dataset and baseline system}
\label{secDataBaseline}

Since there are no publicly available transcribed speech data for Tamil and Kannada, we have collected 150 hours of data for Tamil, and 347 hours for Kannada on our own. The Tamil data was recorded from 531 native Tamil speakers, whereas the Kannada data was recorded from 915 native speakers of Kannada. All the data was recorded in clean, noise-free environment using USB microphones. These two speech corpora have now been made publicly available on OpenSLR \cite{mile_Tamil_asr_data, mile_Kannada_asr_data}. We have also used the 67 hours of Tamil data provided by Microsoft \cite{microsoftdata}. We have divided the data into two parts: (i) training data - 152 hours for Tamil and 275 hours for Kannada and (ii) test data - 65 hours for Tamil and 72 hours for Kannada.

We have developed in-house grapheme-to-phoneme converters for both the languages to phonetize the vocabulary of words in our corpus to build the pronunciation lexicon. We have followed the procedures as explained in \cite{asr_tamil} to build DNN based acoustic models for Tamil and Kannada using Kaldi toolkit \cite{kaldi}. 3-gram language models are estimated using srilm toolkit \cite{srilm} on a large text corpus containing 4.4 million Tamil words and 8 million Kannada words. The language models, lexicon models and the trained acoustic models are combined to form the final decoding graph which is then used for decoding the test data. This word-based ASR system is used as the baseline model for all our experiments.

\begin{figure*}[!ht]
\includegraphics[width=\textwidth,height=0.8\textheight]{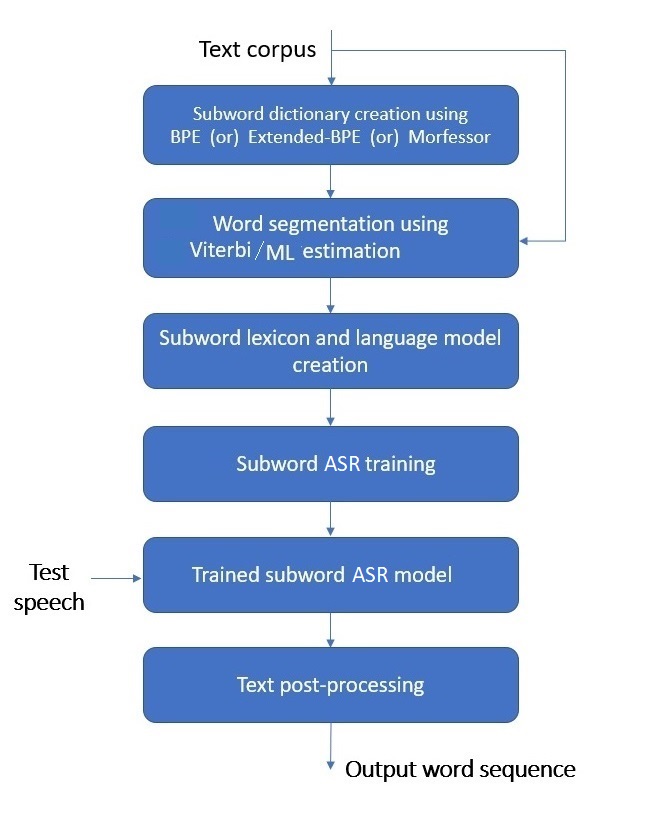}
\centering
\caption{Block diagram for training and testing subword-ASR for Tamil/Kannada.}
	\label{fig_3_automan}
\end{figure*}

The subword-based ASR systems are built using the same procedure except that the words in the transcriptions are segmented to subwords and then given for training. Furthermore, during testing, the output subword sequences from the ASR is post-processed using some deterministic rules to obtain the final word sequence. The block diagram illustrating the training and testing the subword-ASR is shown in Fig. \ref{fig_3_automan}.

\section{Construction of subword dictionary}
\label{sec:subDict}
In this section, we describe different tools and techniques that we have used to automatically construct the subword dictionaries from the given Tamil and Kannada text corpora. We first introduce the well-known technique of BPE \cite{sub_bpe01} and propose a modified version of it called extended-BPE and use each of them independently to create subword dictionary. We then briefly explain the \textit{Morfessor} tool for morphological analysis that we have also used for subword dictionary creation.

\subsection{Subword dictionary construction using BPE} \label{sec:bpe}

BPE is a commonly used technique for data compression \cite{sub_bpe02} which has been extensively used in machine translation, speech recognition and other NLP applications to handle out-of-vocabulary words. BPE constructs the subword dictionary $\mathbb{D}$ from the given text corpus using a bottom-up approach to iteratively build the codebook by combining the most frequently occurring sequence of characters to form a codeword.

The first step is to initialize the codebook with the list of all the characters present in the text corpus. Next, we split the words in the corpus into their constituent characters (separated by whitespaces) and find the most frequently occurring pair of characters in the corpus (say `A B'). We then merge them to form a single codeword (`AB') and add it to the existing codebook and then replace all the occurrences of the codeword sequence `A B' in the corpus by `AB'. These operations are repeated till we reach a desired number of codewords $N$ in the codebook (which is an user-defined hyperparameter). The pseudocode for building the subword dictionary using BPE in a computationally efficient manner is given in Algorithm \ref{alg_3_1}.

\begin{algorithm}
\caption{Subword dictionary construction using byte pair encoding algorithm}
\label{alg_3_1}
\SetAlgoLined
\vspace{4pt}
\KwIn{\hspace{0.3cm} (i) Text corpus \\
      \hspace{1.73cm} (ii) Size of the subword dictionary $N$
}
\vspace{4pt}
\KwOut{
(i) Subword dictionary $\mathbb{D}$ \\
\hspace{1.92cm}(ii) Subword probability mass function $\phi(d): \forall d \in \mathbb{D}$
}
\vspace{4pt}
 Split the lines in the text corpus into words separated by the newline character; \\
 Split all the words into characters separated by whitespaces; \\
 Obtain \textit{n-gram} character counts map functions $\psi_{l}(\cdot): 1 \leq l \leq 7$; \\
 Sort the entries in $\psi_{l}(\cdot)$ in descending order based on their values;\\
 Initialize $\mathbb{D}$ and $\phi(\cdot)$ as empty lists; \\
 \vspace{4pt}
 \For{$i \gets 1$ to length($\psi_{1}(\cdot)$)}{
   $codeword\_char\_seq \gets $ Key of $\psi_{1}(i)$; \\
   $codeword \gets $ Merge the characters in $codeword\_char\_seq$; \\
   $count\_value \gets $ Value of $\psi_{1}(i)$; \\
   Append $codeword$ to $\mathbb{D}$; \\
   Append $count\_value$ to $\phi(\cdot)$; \\
 }
 \vspace{4pt}
 \While{$|\mathbb{D}| < N$}{
  $l^*, i^* \gets \operatorname{argmax}_{\:l,i}  \hspace{0.2cm} \psi_{l}(i)$; \\
  $codeword\_char\_seq \gets $ Key of $\psi_{l^*}(i^*)$; \\
  $codeword \gets $  Merge the characters in $codeword\_char\_seq$; \\
  $count\_value \gets $ Value of $\psi_{l^*}(i^*)$; \\
  Append $codeword$ to $\mathbb{D}$; \\
  Append $count\_value$ to $\phi(\cdot)$; \\
  Delete the entry $\psi_{l^*}(i^*)$; \\
  \vspace{4pt}
  \For{$d \in \mathbb{D}$}{
    \If{($d$ is a substring of $codeword$) and ($\phi(d)$ == $\phi(codeword)$)}{
      Delete the element $d$ in the list $\mathbb{D}$;\\
      Delete the $count\_value$ corresponding to $d$ in the list $\phi(\cdot)$; \\
    }
  }
 }
 \vspace{4pt}
 Normalize $\phi(\cdot)$ such that $\sum_{d\in\mathbb{D}} \phi(d) = 1$;\\
 \textbf{return } $\mathbb{D}$ and $\phi(\cdot)$;
 \vspace{4pt}
\end{algorithm}

\subsection{Subword dictionary creation using extended-BPE} \label{sec:ebpe}
We have customized the  original BPE algorithm so that greedy merging of codeword pairs is avoided to some extent by having adequate number of codewords of different lengths. This is achieved by adding a constraint that there should be at most $N_l$ number of $l$-length codewords in the codebook.

We first split the lines in the given text corpus into word sequences separated by new line characters and then split each word into its constituent characters separated by whitespaces. Next we count the 1-grams, 2-grams, up until 7-grams of characters. We then make a choice of the maximum number ($N_l$) of codewords of length $l$ ($1\leq l \leq 7$) that can be added to the dictionary and initialize the codebook with the 1-gram entries. Then, the most frequently occurring $N_l$ entries of $l$-grams (after merging) are added to the codebook. Necessary conditions are added to delete certain codewords in the dictionary so as to reduce the ambiguity in representing a given word by its constituent subwords during segmentation. Once the dictionary is constructed, the codeword counts are normalized so that we get a probability mass function over the codewords that sums to 1. Algorithm \ref{alg_3_2} gives the pseudocode for efficiently constructing the subword dictionary using the extended-BPE technique.

\begin{algorithm}
\caption{Subword dictionary construction using extended byte pair encoding algorithm}
\label{alg_3_2}
\SetAlgoLined
\KwIn{\hspace{0.1cm} (i) Text corpus \\
      \hspace{1.57cm} (ii) Size of the dictionary $N$ and \{$N_l : 1 \leq l \leq 7$\} such that $\sum_{l}N_l = N$
}
\vspace{2pt}
\KwOut{
(i) Subword dictionary $\mathbb{D}$ \\
\hspace{1.92cm}(ii) Subword probability mass function $\phi(d): \forall d \in \mathbb{D}$
}
 \vspace{2pt}
 Split the lines in the text corpus into words separated by new line character; \\
 Split all the words into characters separated by whitespaces; \\
 Calculate \textit{n-gram} character counts map functions $\psi_{l}(\cdot): 1 \leq l \leq 7$; \\
 Sort the entries in $\psi_{l}(\cdot)$ in descending order based on its values;\\
 Initialize $\mathbb{D}$ and $\phi(\cdot)$ as empty lists; \\
 \vspace{1pt}
 \For{$i \gets 1$ to length($\psi_{1}(\cdot)$)}{
   $codeword\_char\_seq \gets $ Key of $\psi_{1}(i)$; \\
   $codeword \gets$ Merge the characters in $codeword\_char\_seq$; \\
   $count\_value \gets $ Value of $\psi_{1}(i)$; \\
   Append $codeword$ to $\mathbb{D}$; \\
   Append $count\_value$ to $\phi(\cdot)$; \\
 }
 \vspace{1pt}
 \For{$l \gets 2$ to $7$} {
   $n\_dict\_count \gets 0$; \\
   \vspace{1pt}
   \While{$n\_dict\_count < N_l$}{
     $i^* \gets \operatorname{argmax}_{i}  \hspace{0.2cm} \psi_{l}(i)$; \\
     $codeword\_char\_seq \gets $ Key of $\psi_{l}(i^*)$; \\
     $codeword \gets $ Merge the characters in $codeword\_char\_seq$; \\
     $count\_value \gets $ Value of $\psi_{n}(i^*)$; \\
     Append $codeword$ to $\mathbb{D}$; \\
     Append $count\_value$ to $\phi(\cdot)$; \\
     Delete the entry $\psi_{l}(i^*)$; \\
     \vspace{2pt}
     \For{$d \in \mathbb{D}$}{
       \If{($d$ is a substring of $codeword$) and ($\phi(d)$ == $\phi(codeword)$)}{
        Delete the element $d$ in the list $\mathbb{D}$;\\
        Delete the $count\_value$ corresponding to $d$ in the list $\phi(\cdot)$; \\
       }
     }
     \vspace{2pt}
     $n\_dict\_count \gets n\_dict\_count + 1$; \\
   }
 }
 \vspace{2pt}
 Normalize $\phi(\cdot)$ such that $\sum_{d\in\mathbb{D}} \phi(d) = 1$; \\
 \textbf{return } $\mathbb{D}$ and $\phi(\cdot)$;
 \vspace{2pt}
\end{algorithm}

\subsection{Subword dictionary creation using \textit{Morfessor}}   
We have used the \textit{Morfessor} tool to automatically segment words in the vocabulary into their subword sequences \cite{sub_morfessor01}. \textit{Morfessor} is a probabilistic model $\mathcal{M}$ for morphological learning which uses both syntactic and semantic aspects of morphemes that are discovered from the given text corpus. It uses \textit{maximum aposteriori} criterion \cite{sub_morfessor02} to estimate the subword model parameters $\mathcal{M}$ such that $p(\mathcal{M} \; | \; text\_corpus)$ is maximized, i.e.,
\begin{equation}
    \label{eqn_3_1}
    \mathcal{M}^* = argmax_{\mathcal{M}} \;\; p(\mathcal{M} \; | \; text\_corpus)
\end{equation}

The list of unique subwords obtained after segmenting all the words in the text corpus by the \textit{Morfessor} toolkit is used as the subword dictionary $\mathbb{D}$. We then use  $\mathbb{D}$ and apply ML and Viterbi segmentation procedures in an iterative manner, as explained in sections \ref{sec:ml} and \ref{sec:Viterbi}, respectively,  to get the final subword segmented sequence for every word in the vocabulary $\mathbb{V}$.

It is to be noted that the subword dictionary creation using BPE, extended-BPE or Morfessor is computation-driven and not linguistic knowledge-driven. Hence, not all the subwords generated are guaranteed to be like morphemic units.

In the next sections, we explain the ML and Viterbi-based segmentation of words. Word segmentation is a combinatorial search problem which may result in non-unique segmentation. Further, since it is not known as to how many number of segments a given word has to be split into, the segmentation problem becomes more complicated. Since each word in the text corpus needs to be replaced by its corresponding subword sequence, we need to choose only the best (hence unique) sequence of subwords from the list of all possible combinations of subword sequences.

\section{Statistical formulation of maximum likelihood-based word segmentation}   \label{sec:statml}

We now propose a computationally efficient approach based on ML criterion which uses $n$-gram subword language model (which can be implemented in WFST framework as explained in section \ref{sec:ml}) that gives the most-probable subword segments for a given word. The set of subwords $\overline{Z}$ to generate (that concatenates to form) the word $w$ using a given subword dictionary $\mathbb{D}$ is unknown (or a hidden variable). We estimate the optimal subword sequence $\overline{Z}^*$ by maximizing the joint log-likelihood of data and hidden variable $\mathcal{L}(w, \overline{Z}; \theta)$ as given by,
\begin{align}
    \label{eqn_3_2}
    \overline{Z}^* &= argmax_{\overline{Z}} \; \; \mathcal{L}(w, \overline{Z}; \theta) \\[10pt]
    \label{eqn_3_3}
    &= argmax_{\overline{Z}} \; \; \mathcal{L}(w| \overline{Z}; \theta) \; + \; \mathcal{L}(\overline{Z}; \theta) \\[10pt]
   \intertext{\indent where,}
   \label{eqn_3_4}
    \mathcal{L}(w| \overline{Z} ; \theta) &= log(p(w|\overline{Z} ; \theta) \\[10pt]
    \label{eqn_3_5}
    \mathcal{L}(\overline{Z} ; \theta) &= log(p(z_1, z_2, ..., z_S ; \theta)) \\[10pt]
  \intertext{\indent We define $p(w|\overline{Z}; \theta)$ as a Kronecker-delta function and express it as,}
  \label{eqn_3_6}
  p(w| \overline{Z}; \theta) &= \left\{
  \begin{array}{rl}
    1 & : w==concat(\overline{Z})\\
    0 & : otherwise
  \end{array}
  \right. 
   \intertext{where $concat(\overline{Z})$ is concatenation of strings in $\overline{Z}$. $p(\overline{Z})$ can be assumed to be n-gram approximation of the subword sequence. We consider only 1-gram and 2-gram approximations in this derivation. Without loss of generality, this can be extended and derived for any higher order approximations.}
   \label{eqn_3_7}
   \textnormal{For 1-gram case:} \hspace{1cm} p(\overline{Z}; \theta) &\approx \prod_{m=1}^{|\overline{Z}|} \phi(z_m) \\
   \label{eqn_3_8}
   \textnormal{For 2-gram case:} \hspace{1cm} p(\overline{Z}; \theta) &\approx \phi(z_1) \; \; \prod_{m=2}^{|\overline{Z}|} \mathbb{B}(z_m|z_{m-1}) \; \phi(z_m)
\end{align}

where the model parameters $\phi$ and $\mathbb{B}$ (collectively called as $\theta$) are the unigram and bigram probabilities over subwords in the dictionary. To estimate these parameters, we need to maximize the log-likelihood function $\mathcal{L}(w; \; \theta)$ of the given word, which cannot be done directly. Hence we assume some initial values for these parameters, say $\theta_k$ and use expectation-maximization procedure \cite{sub_em} and iteratively maximize the Q-function to find $\theta_{k+1}$:

\begin{align}
    \label{eqn_3_9}
  \theta_{k+1} & = argmax_{\theta} \;\; Q(\theta, \theta_k) \\[10pt]
  \label{eqn_3_10}
  \theta_{k+1} & = argmax_{\theta} \;\; \mathbb{E}_{\:\overline{Z}|w,\theta_k} \left [ \mathcal{L}(w, \overline{Z}; \theta) \right ] 
  \intertext{where,}
  \label{eqn_3_11}
  Q(\theta, \theta_k) & = \mathbb{E}_{\:\overline{Z}|w,\theta_k} \left [  \mathcal{L}(w, \overline{Z}; \theta) \right ] \\[10pt]
  \label{eqn_3_12}
  & = \mathbb{E}_{\:\overline{Z}|w,\theta_k} \left [ log \left ( p(w, \overline{Z}; \theta) \right) \right ] \\[10pt]
  \label{eqn_3_13}
  & = \mathbb{E}_{\:\overline{Z}|w,\theta_k}  \left [ log \left ( p(w|\overline{Z}; \theta) \prod_{m=1}^{|\overline{Z}|} \phi(z_m) \; \prod_{m=2}^{|\overline{Z}|} \mathbb{B}(z_m | z_{m-1}) \right ) \right ] \\[10pt]
  \label{eqn_3_14}
  & = \mathbb{E}_{\:\overline{Z}|w,\theta_k} \left [  log(p(w|\overline{Z}; \theta)) + \sum_{m=1}^{|\overline{Z}|} log(\phi(z_m)) \; + \; \sum_{m=2}^{|\overline{Z}|} log(\mathbb{B}(z_m | z_{m-1})) \right ] \\[10pt]
  \label{eqn_3_14b}
  \begin{split}
  & = \sum_{\overline{Z} \in \mathbb{W}_w} p(\overline{Z}|w;\theta_k) \left ( log(p(w|\overline{Z};\theta) \right) + \sum_{\overline{Z} \in \mathbb{W}_w} p(\overline{Z}|w;\theta_k) \left (\sum_{m=1}^{|\overline{Z}|} log(\phi(z_m)) \right ) \\
  & \qquad + \;  \sum_{\overline{Z} \in \mathbb{W}_w} p(\overline{Z}|w;\theta_k) \left ( \sum_{m=2}^{|\overline{Z}|} log(\mathbb{B}(z_m | z_{m-1})) \right )
  \end{split} \\[10pt]
  \label{eqn_3_15}
  \begin{split}
  & = \sum_{\overline{Z} \in \mathbb{W}_w} p(\overline{Z}|w;\theta_k) \left (\sum_{m=1}^{|\overline{Z}|} log(\phi(z_m)) \right ) \\
  & \qquad + \;  \sum_{\overline{Z} \in \mathbb{W}_w} p(\overline{Z}|w;\theta_k) \left ( \sum_{m=2}^{|\overline{Z}|} log(\mathbb{B}(z_m | z_{m-1})) \right )
  \end{split} \\[10pt]
  \label{eqn_3_16}
  & = Q_1(\phi, \phi_k) + Q_2(\mathbb{B}, \mathbb{B}_k)
\end{align}

where $\mathbb{W}_w$ is a set where each element is a sequence of subwords such that the concatenated sequence forms the word $w$. By the definition of $p(w|\overline{Z})$ in equation \ref{eqn_3_6}, the expectation over all possible $\overline{Z}$ reduces to expectation over only those $\overline{Z} \in \mathbb{W}_w$, thus eliminating the first term $p(w|\overline{Z};\theta)$ in equation \ref{eqn_3_14b} (since it becomes zero). The posterior term $p(\overline{Z}|w;\theta_k)$ is defined as $\gamma_{k}(\overline{Z})$ and can be calculated by,

\begin{equation}
    \label{eqn_3_17}
  \gamma_{k}(\overline{Z}) = \frac{\prod_{j=1}^{|\overline{Z}|} \phi_k(z_j) \mathbb{B}_k(z_{j}|z_{j-1})}{\sum_{\overline{Z} \in \mathbb{W}_w} \prod_{j=1}^{|{z}|} \phi_k(z_j) \mathbb{B}_k(z_{j}|z_{j-1})}
\end{equation}

The new estimates of $\phi_{k+1}$ and $\mathbb{B}_{k+1}$ are obtained by independently maximizing the functions $Q_1(\phi, \phi_k)$ and $Q_2(\mathbb{B}, \mathbb{B}_k)$ respectively, where,

\begin{align}
    \label{eqn_3_18}
    Q_1(\phi, \phi_k) & = \sum_{\overline{Z} \in \mathbb{W}_w} \gamma_{k}(\overline{Z}) \left (\sum_{m=1}^{|\overline{Z}|} log(\phi(z_m)) \right ) \\[10pt]
    \label{eqn_3_19}
    Q_2(\mathbb{B}, \mathbb{B}_k) &= \sum_{\overline{Z} \in \mathbb{W}_w} \gamma_{k}(\overline{Z}) \left ( \sum_{m=2}^{|\overline{Z}|} log(\mathbb{B}(z_m | z_{m-1})) \right )
\end{align}

\noindent
\textbf{Estimation of $\phi$ : } We estimate $\phi_{k+1}(z_l)$ by maximizing the objective function $Q_1(\phi, \phi_k)$ with the constraint $\sum_{i = 1}^{|\mathbb{D}|} \phi(z_i) = 1$. The Lagrangian function $\mathcal{L}_1$ for this optimization with Lagrangian multiplier $\lambda$ is given by,

\begin{align}
    \label{eqn_3_20}
  \mathcal{L}_1 & = Q_1(\phi, \phi_k) + \lambda \left (\sum_{i = 1}^{|\mathbb{D}|} \phi(z_i) - 1 \right ) 
  \intertext{Next, we set the derivative of $\mathcal{L}_1$ to 0 to solve for $\phi_{k+1}(z_l)$.}
  \label{eqn_3_21}
  \frac{\partial \mathcal{L}_1}{\partial \phi(z_l)} & = \frac{\partial}{\partial \phi(z_l)} \left [ Q_1(\phi, \phi_k) + \lambda (\sum_{i = 1}^{|\mathbb{D}|} \phi(z_i) - 1) \right ] \\[10pt]
  \label{eqn_3_22}
  & = \frac{\partial}{\partial \phi(z_l)} \left [ \left ( \sum_{\overline{Z} \in \mathbb{W}_w} \gamma_{k}(\overline{Z}) \sum_{m=1}^{|\overline{Z}|} log(\phi(z_m)) \right) + \lambda \left ( \sum_{i = 1}^{|\mathbb{D}|} \phi(z_{i}) - 1\right ) \right ]
\end{align}

We define \textit{counts} function $\mathbb{C}(z \textnormal{ in } \overline{Z})$ as the count of the number of times the subword $z$ occurs in the sequence $\overline{Z}$. Setting $\frac{\partial L_1}{\partial \phi(z_l)} = 0$, and solving for $\phi_{k+1}$, we get,

\begin{align}
    \label{eqn_3_23}
  \sum_{\overline{Z} \in \mathbb{W}_w}  \frac{\gamma_{k}(\overline{Z}) \; \mathbb{C}(z_l \textnormal{ in } \overline{Z})}{\phi_{k+1}(z_l)} + \lambda & = 0 \\[10pt]
  \label{eqn_3_24}
  \sum_{\overline{Z} \in \mathbb{W}_w} \gamma_{k}(\overline{Z}) \; \mathbb{C}(z_l \textnormal{ in } \overline{Z}) + \lambda \; \phi_{k+1}(z_l) & = 0 \\[10pt]
  \intertext{\indent Summing equation \ref{eqn_3_24} over index $l$, we get,}
  \label{eqn_3_25}
  \sum_{l = 1}^{|\mathbb{D}|} \sum_{\overline{Z} \in \mathbb{W}_w} \gamma_{k}(\overline{Z}) \; \mathbb{C}(z_l \textnormal{ in } \overline{Z}) + \lambda \sum_{l = 1}^{|\mathbb{D}|} \phi_{k+1}(z_l) & = 0 \\[10pt]
  \intertext{\indent Substituting the constraint $\sum_{i = 1}^{|\mathbb{D}|} \phi(z_i) = 1$ in equation \ref{eqn_3_25}, we get,}
  \label{eqn_3_26}
  \sum_{l = 1}^{|\mathbb{D}|} \sum_{\overline{Z} \in \mathbb{W}_w} \gamma_{k}(\overline{Z}) \; \mathbb{C}(z_l \textnormal{ in } \overline{Z}) + \lambda  & = 0
\end{align}

\begin{align}
    \label{eqn_3_27}
    \lambda  & = - \sum_{l = 1}^{|\mathbb{D}|} \sum_{\overline{Z} \in \mathbb{W}_w} \gamma_{k}(\overline{Z}) \; \mathbb{C}(z_l \textnormal{ in } \overline{Z})
\end{align}

Substituting equation \ref{eqn_3_27} in equation \ref{eqn_3_24}, we get,

\begin{equation}
\label{eqn_3_28}
\boxed{\phi_{k+1}(z_l) = \frac{\sum_{\overline{Z} \in \mathbb{W}_w} \gamma_{k}(\overline{Z}) \; \mathbb{C}(z_l \textnormal{ in } \overline{Z})} {\sum_{l' = 1}^{|\mathbb{D}|} \sum_{\overline{Z} \in \mathbb{W}_w} \gamma_{k}(\overline{Z}) \; \mathbb{C}(z_{l'} \textnormal{ in } \overline{Z})}}
\end{equation}

\vspace{0.8cm}
\noindent
\textbf{Estimating $\mathbb{B}$:} We estimate $\mathbb{B}_{k+1}(z_l|z_m)$ by maximizing $Q_2(\mathbb{B}, \mathbb{B}_k)$ with the constraint $\sum_{i = 1}^{|\mathbb{D}|} \mathbb{B}(z_{i}|z_{j}) = 1$ for any $1 \leq j \leq |\mathbb{D}|$. The Lagrangian function $\mathcal{L}_2$ for this optimization problem with Lagrangian multiplier $\lambda$ is given by,

\begin{align}
    \label{eqn_3_29}
  \mathcal{L}_2 & = Q_2(\mathbb{B}, \mathbb{B}_k) + \lambda \left (\sum_{i = 1}^{|\mathbb{D}|} \mathbb{B}(z_{i}|z_{j}) - 1 \right ) \\[5pt]
  \intertext{\indent The partial derivative of the Lagrangian function $\mathcal{L}_2$ with respect to $\mathbb{B}(z_l | z_m)$ is given by,}
  \label{eqn_3_30}
  \frac{\partial \mathcal{L}_2}{\partial \mathbb{B}(z_l | z_m)} & = \frac{\partial}{\partial \mathbb{B}(z_l|z_m)} \left [ Q_2(\mathbb{B}, \mathbb{B}_k) \; + \; \lambda \left ( \sum_{i = 1}^{|\mathbb{D}|} \mathbb{B}(z_i | z_j) - 1) \right ) \right ]
\end{align}

\begin{align}
    \label{eqn_3_31}
  & = \frac{\partial}{\partial \mathbb{B}(z_l|z_m)} \left [ \sum_{\overline{Z} \in \mathbb{W}_w} \sum_{m' = 2}^{|\overline{Z}|} \gamma_k(\overline{Z}) \: log(\mathbb{B}(z_{m'} | z_{m'-1})) \; + \; \lambda \left ( \sum_{i = 1}^{|\mathbb{D}|} \mathbb{B}(z_i | z_j) - 1 \right ) \right ]
\end{align}

Setting the above partial derivative expression to 0 and solving for $\mathbb{B}_{k+1}(z_l|z_m)$,

\begin{align}
    \label{eqn_3_32}
  \sum_{\overline{Z} \in \mathbb{W}_w} \frac{\gamma_{k}(\overline{Z}) \mathbb{C}(z_{l}z_{m} \textnormal{ in } \overline{Z})}{\mathbb{B}_{k+1}(z_{l}|z_{m})} \; + \; \lambda & = 0 \\[10pt]
  \intertext{\indent where the \textit{counts} function $\mathbb{C}(z_{l}z_{m} \textnormal{ in } \overline{Z})$ is the count of the number of times the strings $z_l$ and $z_m$ occur together as a sequence in $\overline{Z}$}.
  \label{eqn_3_33}
  \sum_{\overline{Z} \in \mathbb{W}_w} \gamma_{k}(\overline{Z}) \mathbb{C}(z_{l}z_{m} \textnormal{ in } \overline{Z}) \; + \; \lambda \mathbb{B}_{k+1}(z_{l}|z_{m}) & = 0 \\[10pt]
  \intertext{Summing the above over index $l$ and using the constraint $\sum_{i = 1}^{|\mathbb{D}|} \mathbb{B}(z_{i}|z_{j}) = 1$,}
  \label{eqn_3_34}
  \sum_{l = 1}^{|\mathbb{D}|} \sum_{\overline{Z} \in \mathbb{W}_w} \gamma_{k}(\overline{Z}) \mathbb{C}(z_{l}z_{m} \textnormal{ in } \overline{Z}) \; + \; \lambda \sum_{l = 1}^{|\mathbb{D}|} \mathbb{B}_{k+1}(z_{l}|z_{m}) & = 0 \\[10pt]
  \label{eqn_3_35}
  \sum_{l = 1}^{|\mathbb{D}|} \sum_{\overline{Z} \in \mathbb{W}_w} \gamma_{k}(\overline{Z}) \mathbb{C}(z_{l}z_{m} \textnormal{ in } \overline{Z}) \; + \; \lambda & = 0
\end{align}

\begin{equation}
    \label{eqn_3_36}
    \lambda = - \sum_{l = 1}^{|\mathbb{D}|} \sum_{\overline{Z} \in \mathbb{W}_w} \gamma_{k}(\overline{Z}) \mathbb{C}(z_{l}z_{m} \textnormal{ in } \overline{Z})
\end{equation}

Substituting the expression for $\lambda$ in equation \ref{eqn_3_33}, we get,

\begin{equation}
\label{eqn_3_37}
\hspace{-1.4cm}
\boxed{\mathbb{B}_{k+1}(z_{l}|z_{m}) = \frac{\sum_{\overline{Z} \in \mathbb{W}_w} \gamma_{k}(\overline{Z}) \mathbb{C}(z_{l}z_{m} \textnormal{ in } \overline{Z})}{\sum_{l' = 1}^{|\mathbb{D}|} \sum_{\overline{Z} \in \mathbb{W}_w} \gamma_{k}(\overline{Z}) \mathbb{C}(z_{l'}z_{m} \textnormal{ in } \overline{Z})}}
\end{equation}

Equations \ref{eqn_3_28} and \ref{eqn_3_37} give the expressions for estimating the model parameters $\phi$ and $\mathbb{B}$ if we consider only one word $w$ in the word vocabulary $\mathbb{V}$ at a time. When we consider all the words in $\mathbb{V}$ at the same time to estimate the parameters, equations \ref{eqn_3_28} and \ref{eqn_3_37} modify to,

\begin{empheq}[box=\fbox]{align}
    \label{eqn_3_38}
  \phi_{k+1}(z_l) & = \frac{\sum_{w \in \mathbb{V}} \sum_{\overline{Z} \in \mathbb{W}_w} \gamma_{k}(\overline{Z}) \; \mathbb{C}(z_l \textnormal{ in } \overline{Z})} {\sum_{w \in \mathbb{V}} \sum_{l' = 1}^{|\mathbb{D}|} \sum_{\overline{Z} \in \mathbb{W}_w} \gamma_{k}(\overline{Z}) \; \mathbb{C}(z_{l'} \textnormal{ in } \overline{Z})} \\[10pt]
  \label{eqn_3_39}
  \mathbb{B}_{k+1}(z_{l}|z_{m}) & = \frac{\sum_{w \in \mathbb{V}} \sum_{\overline{Z} \in \mathbb{W}_w} \gamma_{k}(\overline{Z}) \mathbb{C}(z_{l}z_{m} \textnormal{ in } \overline{Z})}{\sum_{w \in \mathbb{V}} \sum_{l' = 1}^{|\mathbb{D}|} \sum_{\overline{Z} \in \mathbb{W}_w} \gamma_{k}(\overline{Z}) \mathbb{C}(z_{l'}z_{m} \textnormal{ in } \overline{Z})}
\end{empheq}

Finally, using equations \ref{eqn_3_38} and \ref{eqn_3_39}, we estimate the parameters for ${(k+1)}^{th}$ iteration by considering all the words in the vocabulary $\mathbb{V}$ at the same time. Then, keeping them as the current estimates, we calculate the posteriors and reestimate the new parameters for ${(k+2)}^{th}$ iteration. This process is repeated for 15 iterations, and at every iteration, the log-likelihood of the data is guaranteed to increase. The parameters obtained after the final iteration (say $\theta^*$) are used to segment any given $w$ to get the optimal subword sequence $\overline{Z}^*$ using the equation below,

\begin{equation}
    \label{eqn_3_40}
  \overline{Z}^* = argmax_{\overline{Z}} \; \; \mathcal{L}(w|\overline{Z}; \theta^*) \; + \; \mathcal{L}(\overline{Z}; \theta^*);
\end{equation}


\section{Statistical formulation of Viterbi-based word segmentation}
\label{sec:statvit}

The Viterbi-based word segmentation takes $max_{\: \overline{Z}|w;\theta_k} \left [ \; \cdot \; \right]$ operation on the joint likelihood function $\mathcal{L}(w,\overline{Z}; \theta)$ to estimate the model parameters unlike  ML-based approach that takes expectation $\mathbb{E}_{\: \overline{Z}|w;\theta_k} \left [ \; \cdot \; \right ]$ operation.  The joint log-likelihood function is given by,

\begin{align}
    \label{eqn_3_41}
    \mathcal{L}(w,\overline{Z}; \theta) & = log(p(w, \overline{Z}; \theta) \\[5pt]
\intertext{The parameters $\theta$ are obtained by maximizing the objective function:}
    \label{eqn_3_42}
    Q(\theta, \theta_k) & = max_{\overline{Z}|w;\theta_k} \;\; \mathcal{L}(w,\overline{Z}; \theta) \\[10pt]
    \label{eqn_3_43}
    & = max_{\overline{Z}|w;\theta_k} \;\; log(p(w, \overline{Z}; \theta)) \\[10pt]
    \intertext{\indent Substituting the expression for $p(w, \overline{Z}; \theta)$ given by equations \ref{eqn_3_12} and \ref{eqn_3_13} (as per ML formulation), we can write the above equation as,}
    \label{eqn_3_44}
    & = max_{\overline{Z}|w;\theta_k} \; \left [ log \left( p(w|\overline{Z}; \theta) \prod_{m=1}^{|\overline{Z}|} \phi(z_{m}) \;  \prod_{m=2}^{|\overline{Z}|} \mathbb{B}(z_{m} | z_{m-1}) \right ) \right ] \\[10pt]
    \label{eqn_3_45}
    \begin{split}
    & = max_{\overline{Z}|w;\theta_k} \; \left [ log(p(w|\overline{Z}; \theta)) + log \left( \prod_{m=1}^{|\overline{Z}|} \phi(z_{m}) \right ) \right ] \\[5pt]
    & \qquad + max_{\overline{Z}|w;\theta_k} \; \left [ log \left ( \prod_{m=2}^{|\overline{Z}|} \mathbb{B}(z_{m} | z_{m-1}) \right ) \right ]
    \end{split} \\[10pt]
    \label{eqn_3_46}
    \begin{split}
    & = max_{\:\overline{Z} \in \mathbb{W}_w} \; \left [ log \left( \prod_{m=1}^{|\overline{Z}|} \phi(z_{m}) \right ) \right ] \\[5pt]
    & \qquad + \;  max_{\:\overline{Z} \in \mathbb{W}_w} \; \left [ log \left ( \prod_{m=2}^{|\overline{Z}|} \mathbb{B}(z_{m} | z_{m-1}) \right ) \right ]
    \end{split} \\[5pt]
    \label{eqn_3_47}
    & = Q_{1}(\phi, \phi_k) + Q_{2}(\mathbb{B}, \mathbb{B}_k) \\[5pt]
\intertext{where $\mathbb{W}_w$ contains the set of subword sequences, where each sequence when concatenated forms the word $w$, and $Q_{1}(\phi, \phi_k)$ and $Q_{2}(\mathbb{B}, \mathbb{B}_k)$ are given by,}
    \label{eqn_3_48}
    Q_{1}(\phi, \phi_k) & = max_{\:\overline{Z} \in \mathbb{W}_w} \; \left [ log \left( \prod_{m=1}^{|\overline{Z}|} \phi(z_{m}) \right ) \right ] \\[10pt]
    \label{eqn_3_49}
    Q_{2}(\mathbb{B}, \mathbb{B}_k) & = max_{\:\overline{Z} \in \mathbb{W}_w} \; \left [ log \left ( \prod_{m=2}^{|\overline{Z}|} \mathbb{B}(z_{m} | z_{m-1}) \right ) \right ]
\end{align}

\noindent
\textbf{Estimation of $\phi$ :} By maximizing $Q_{1}(\phi, \phi_k)$ with the constraint $\sum_{i=1}^{|\mathbb{D}|}\phi(z_{i}) = 1$, we obtain $\phi_{k+1}(z_l)$. The Lagrangian function $\mathcal{L}_1$ is given by,
\begin{align}
    \label{eqn_3_50}
    \mathcal{L}_1 = max_{\:\overline{Z} \in \mathbb{W}_w} \left [ log \left( \prod_{m=1}^{|\overline{Z}|} \phi(z_{m}) \right ) \right ] + \lambda \left ( \sum_{m=1}^{|\mathbb{D}|} \phi(z_{m}) - 1 \right )
\end{align}

Taking partial derivative of $\mathcal{L}_1$ w.r.t. $\phi_(z_m)$ and setting it to 0 to solve for $\phi_{k+1}(z_m)$, we get,

\begin{equation}
    \label{eqn_3_51}
   \frac{\partial}{\partial \phi(z_m)} \left [ max_{\overline{Z}|w; \theta_k} \left [ \sum_{m=1}^{|\overline{Z}|} log(\phi(z_{m})) \right ] \; + \; \lambda \left ( \sum_{m = 1}^{|\mathbb{D}|} \phi_{k+1}(z_{m}) - 1\right ) \right ] = 0
\end{equation}

Let $\overline{Z}^*$ be the sequence which maximizes the first summation term in equation \ref{eqn_3_51}. This equation now reduces to,
\begin{align}
    \label{eqn_3_52}
   \frac{\mathbb{C}(z_m \textnormal{ in } \overline{Z}^*)}{\phi_{k+1}(z_m)} \; + \; \lambda &= 0 \\[10pt]
   \label{eqn_3_53}
   \mathbb{C}(z_m \textnormal{ in } \overline{Z}^*) \; + \; \lambda \phi_{k+1}(z_m) &= 0 \\
   \intertext{\indent Summing over $m$ and using the constraint, we get,}
   \label{eqn_3_55}
   \sum_{m = 1}^{|\mathbb{D}|} \mathbb{C}(z_m \textnormal{ in } \overline{Z}^*) \; + \; \lambda \sum_{m = 1}^{|\mathbb{D}|} \phi_{k+1}(z_m) & = 0 \\[10pt]
   \label{eqn_3_56}
   \sum_{m = 1}^{|\mathbb{D}|} \mathbb{C}(z_m \textnormal{ in } \overline{Z}^*) \; + \; \lambda & = 0 \\[10pt]
   \label{eqn_3_57}
   \lambda & = - \sum_{m = 1}^{|\mathbb{D}|} \mathbb{C}(z_m \textnormal{ in } \overline{Z}^*)
\end{align}

Substituting the above expression for $\lambda$ in equation \ref{eqn_3_53}, we get the final estimate for $\phi_{k+1}(z_m)$:

\begin{equation}
\label{eqn_3_58}
\boxed{
   \phi_{k+1}(z_m) = \frac{\mathbb{C}(z_m \textnormal{ in } \overline{Z}^*) } { \sum_{m'=1}^{|\mathbb{D}|} \mathbb{C}(z_{m^{'}} \textnormal{ in } \overline{Z}^*) } }
\end{equation}

\noindent
\textbf{Estimation of $\mathbb{B}$:} To estimate $\mathbb{B}$, we maximize $Q_{2}(\mathbb{B}, \mathbb{B}_k)$ with the constraint $\sum_{i = 1}^{|\mathbb{D}|} \mathbb{B}(z_{i}|z_{j}) = 1$ for any $1 \leq j \leq |\mathbb{D}|$. The Lagrangian function $\mathcal{L}_2$ is written as,

\begin{align}
    \label{eqn_3_59}
    \mathcal{L}_2 & = max_{\overline{Z} | w; \theta_k} \left [ \sum_{l'=1}^{|\overline{Z}|} log(\mathbb{B}(z_{l'}|z_{m'})) \right ] + \lambda \left ( \sum_{l'=1}^{|\mathbb{D}|} \mathbb{B}(z_{l'}|z_{m'}) - 1 \right )
\end{align}

Taking partial derivative of $\mathcal{L}_2$ with respect to $\mathbb{B}(z_l | z_m)$ and setting it to 0 to solve for $\mathbb{B}_{k+1}(z_l | z_m)$, we get,

\begin{align}
    \label{eqn_3_60}
   \frac{\partial \mathcal{L}_2}{\partial \mathbb{B}(z_l|z_m)} & = 0
\end{align}

\begin{equation}
    \label{eqn_3_61}
    \frac{\partial} {\partial \mathbb{B}(z_l|z_m)} \left[ max_{\overline{Z} | w; \theta_k} \left [ \sum_{l'=1}^{|\overline{Z}|} log(\mathbb{B}(z_{l'}|z_{m})) \right ] + \lambda \left ( \sum_{i=1}^{|\mathbb{D}|} \mathbb{B}(z_{i}|z_{j}) - 1 \right ) \right ] = 0
\end{equation}

Assuming $\overline{Z}^*$ to be the subword sequence that maximizes the first summation term in the above equation, we get,

\begin{align}
    \label{eqn_3_62}
   \frac{\mathbb{C}(z_l z_m \textnormal{ in } \overline{Z}^* )} {\mathbb{B}_{k+1}(z_l|z_m)} \; + \; \lambda & = 0 \\[5pt]
   \label{eqn_3_63}
   \mathbb{C}(z_l z_m \textnormal{ in } \overline{Z}^* ) \; + \; \lambda \; \mathbb{B}_{k+1}(z_l|z_m) & = 0 \\[5pt]
   \intertext{\indent Summing the above equation over the index $l$ and using the constraint $\sum_{i = 1}^{|\mathbb{D}|} \mathbb{B}(z_{i}|z_{j}) = 1$, we get,}
   \label{eqn_3_64}
   \sum_{l=1}^{|\mathbb{D}|} \mathbb{C}(z_l z_m \textnormal{ in } \overline{Z}^* ) \; + \; \lambda \; \sum_{l=1}^{|\mathbb{D}|}\mathbb{B}_{k+1}(z_l|z_m) & = 0
\end{align}

\begin{align}
    \label{eqn_3_65}
   \sum_{l=1}^{|\mathbb{D}|} \mathbb{C}(z_l z_m \textnormal{ in } \overline{Z}^* ) \; + \; \lambda & = 0 \\[10pt]
   \label{eqn_3_66}
   \lambda & = - \sum_{l=1}^{|\mathbb{D}|} \mathbb{C}(z_l z_m \textnormal{ in } \overline{Z}^* )
\end{align}

Substituting equation \ref{eqn_3_66} in \ref{eqn_3_63}, we get the estimate for $\mathbb{B}_{k+1}(z_l|z_m)$ as,

\begin{equation}
\label{eqn_3_67}
\boxed{
   \mathbb{B}_{k+1}(z_l|z_m) = \frac{\mathbb{C}(z_l z_m \textnormal{ in } \overline{Z}^* )} {\sum_{l'=1}^{|\mathbb{D}|} \mathbb{C}(z_{l^{'}} z_m \textnormal{ in } \overline{Z}^* )} }
\end{equation}

\vspace{0.5cm}
Unlike ML, Viterbi segmentation requires $\overline{Z}^*$ to calculate the model parameters $\phi$ and $\mathbb{B}$. Thus, before maximization step, we need to get $\overline{Z}^*$ using equation \ref{eqn_3_68} and use it in equations \ref{eqn_3_58} and \ref{eqn_3_67}.

\begin{align}
    \label{eqn_3_68}
    \overline{Z}^* = argmax_{\overline{Z}|w; \theta_k} \; \; \mathcal{L}(w,\overline{Z}; \theta)
\end{align}

Equations \ref{eqn_3_58} and \ref{eqn_3_67} are the parameter update equations when we consider only one word in the vocabulary $\mathbb{V}$ at a time. If we consider all the words in $\mathbb{V}$ at the same time, then these equations modify to,

\begin{empheq}[box=\fbox]{align}
\label{eqn_69}
  \phi_{k+1}(z_m) & = \frac{\sum_{w \in \mathbb{V}}\mathbb{C}(z_m \textnormal{ in } \overline{Z}_{w}^{*}) } { \sum_{w \in \mathbb{V}} \sum_{m'=1}^{|\mathbb{D}|} \mathbb{C}(z_{m^{'}} \textnormal{ in } \overline{Z}_{w}^{*}) } \\[10pt]
  \label{eqn_70}
  \mathbb{B}_{k+1}(z_l|z_m) & = \frac{\sum_{w \in \mathbb{V}} \mathbb{C}(z_l z_m \textnormal{ in } \overline{Z}_{w}^{*} )} {\sum_{w \in \mathbb{V}} \sum_{l'=1}^{|\mathbb{D}|} \mathbb{C}(z_{l^{'}} z_m \textnormal{ in } \overline{Z}_{w}^{*} )}
\end{empheq}

where $\overline{Z}_{w}^{*}$ is the optimal segmented subword sequence for the word $w$.

\section{WFST implementation of ML word segmentation}   \label{sec:ml}
The parameter update equations \ref{eqn_3_38} and \ref{eqn_3_39} for ML segmentation require us to compute the set $\mathbb{W}_w$ for every word $w$ in the vocabulary, which contains the list of all possible subword sequences $\overline{Z}$ that concatenate to form the word $w$. Calculating the set $\mathbb{W}_w$ is a computationally expensive task, and so we have developed a technique which efficiently computes all sequences and their likelihoods for a given word using WFST framework \cite{sub_ml01}.

Given the subword dictionary $\mathbb{D}$, its unigram $\phi$ and bigram probability densities $\mathbb{B}$, and the list of words in the vocabulary $\mathbb{V}$, we create three WFSTs namely subword dictionary WFST (SD-WFST), subword grammar WFST (SG-WFST) and word WFSTs (W-WFST) in a specific manner and compose them together to obtain O-WFST and perform a simple search operation on it to get the list of all possible subword sequences $\mathbb{W}_w$ (and their corresponding likelihoods) that form the given word $w$. These likelihood values are then used as posteriors $\gamma_k(\overline{Z_w})$ in the equations \ref{eqn_3_38} and \ref{eqn_3_39}. $\mathbb{C}(z_l \textnormal{ in } \overline{Z}_w)$ is calculated by counting the number of occurrences of any subword $z_l$ in the segmented subword sequence $\overline{Z}_w$. To calculate $\overline{Z}^*$ in equations \ref{eqn_3_40}, we simply choose the subword sequence that corresponds to the maximum weight.

\subsection*{\textbf{Construction of SD-WFST :}} 
\begin{figure*}[!ht]
\includegraphics[width=\textwidth,height=0.5\textheight]{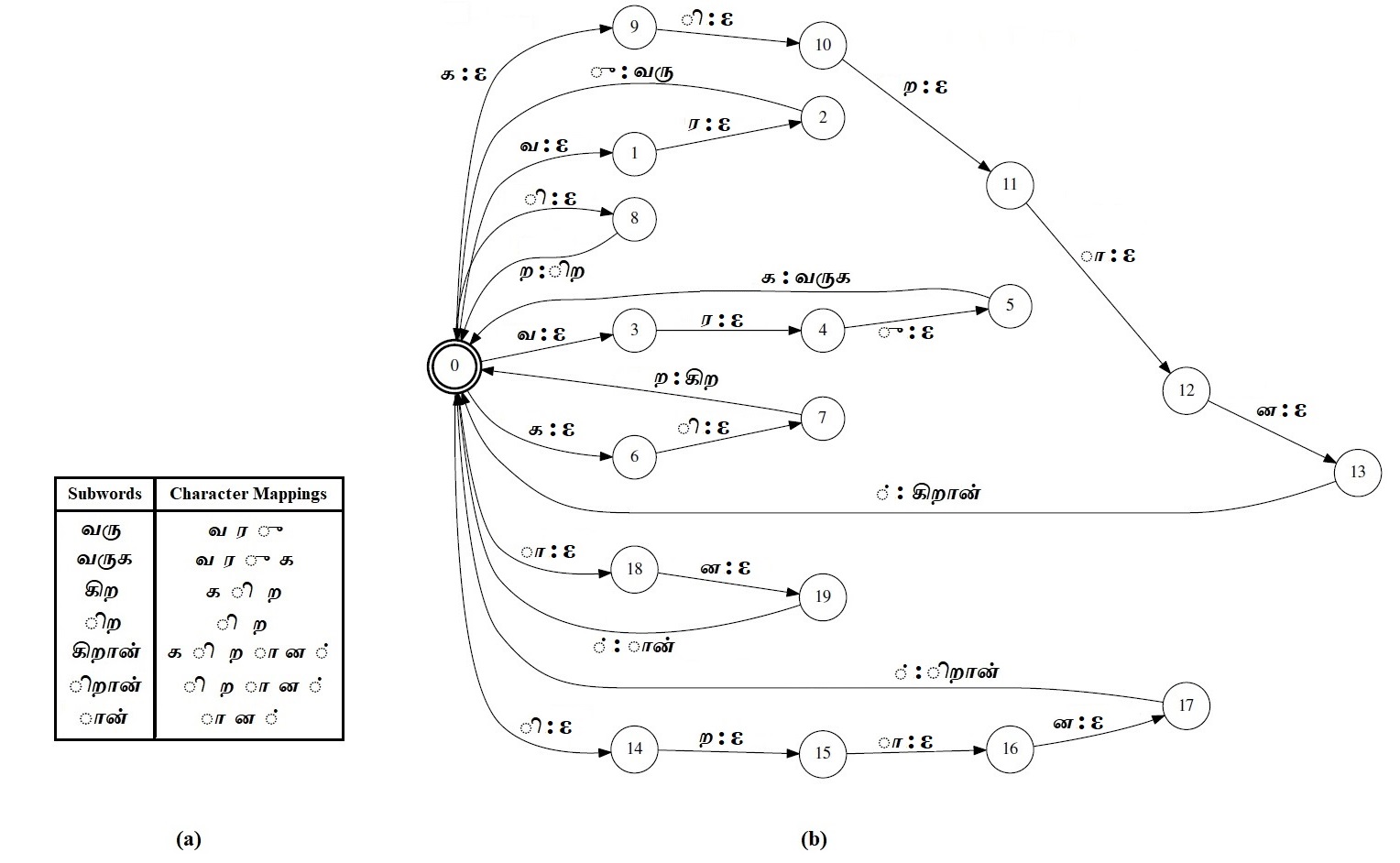}
\centering
\caption{(a) An example subword dictionary $\mathbb{D}$, (b) Subword dictionary weighted finite state transducer (SD-WFST) graph created for $\mathbb{D}$ for Tamil language}
	\label{fig_3_1}
\end{figure*}

\begin{figure*}[!ht]
\centering
\includegraphics[scale= 0.45]{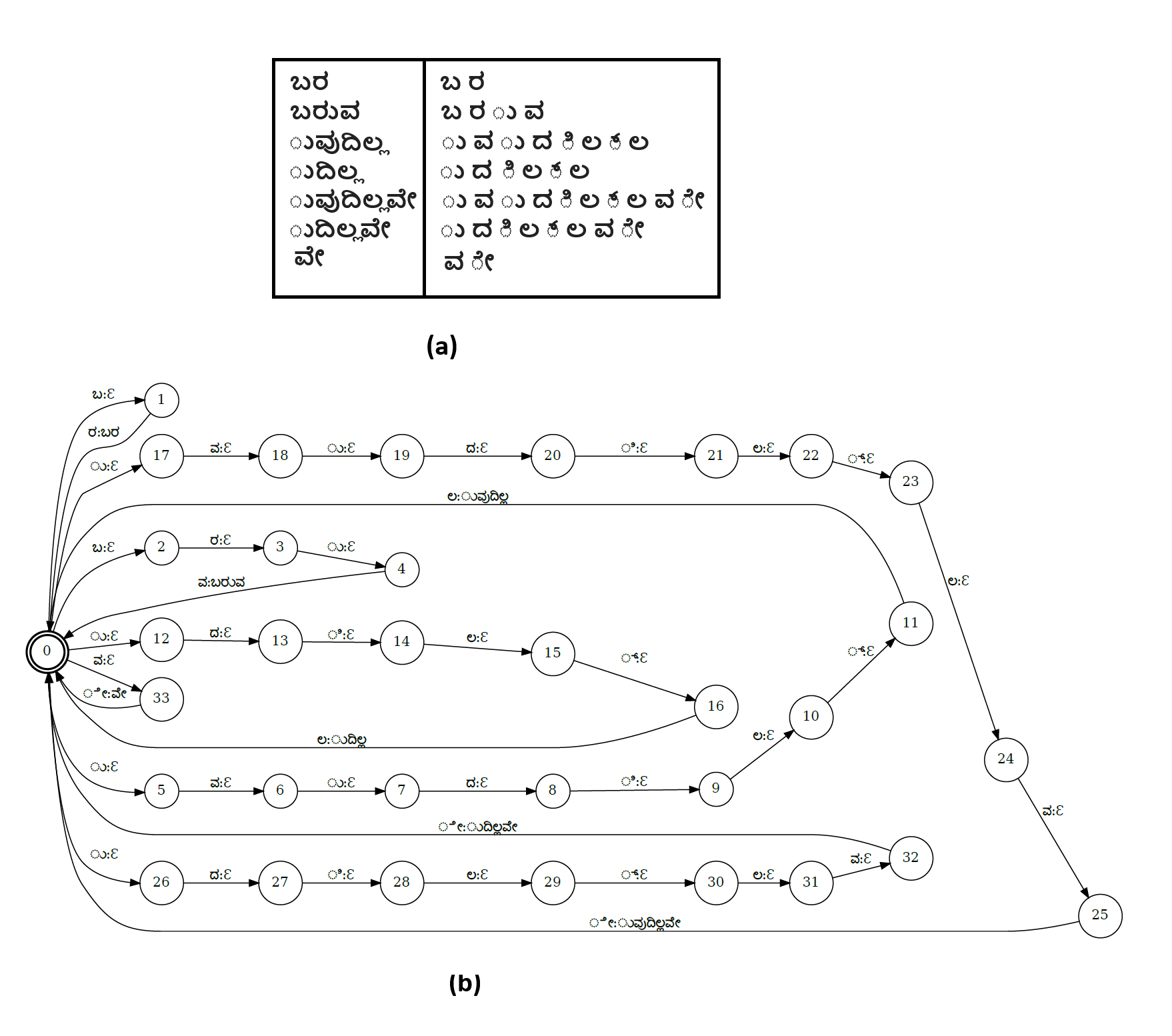}
\caption{ (a) An example subword dictionary $\mathbb{D}$, (b) Subword dictionary weighted finite state transducer graph created for $\mathbb{D}$ for Kannada language}
\label{sd_wfst}
\end{figure*}

We construct this by creating one path for each entry in the subword dictionary. All the paths have empty weights (probability 1.0) and they start and end in the same state (called loop state). The input labels for the transitions/arcs in a path are the individual characters in the subword and the output labels are $\epsilon$ for all the transitions except for the last transition which is the subword string encoded by that path. Figures \ref{fig_3_1} and  \ref{sd_wfst} show a SD-WFST graph for an example dictionary $\mathbb{D}$ for Tamil and Kannada respectively.

\subsection*{\textbf{Construction of SG-WFST :}} 
\begin{figure*}[!ht]
\centering
\includegraphics[width=\textwidth,height=0.5\textheight]{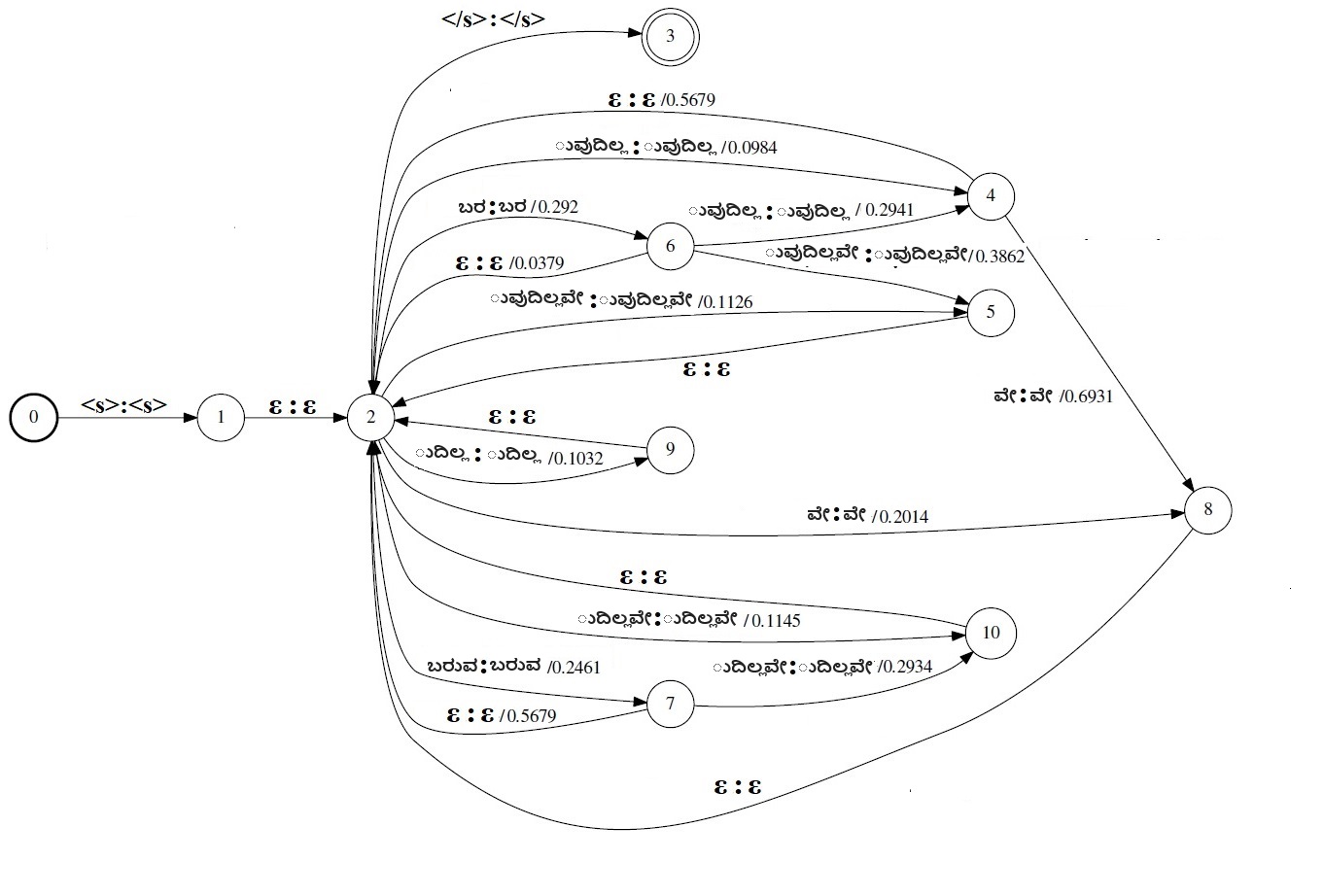}
\caption{A sample Kannada subword grammar weighted finite state transducer (SG-WFST) generated for the subword dictionary $\mathbb{D}$ shown in figure \ref{sd_wfst}a}
\label{sg_wfst}
\end{figure*}

We construct subword grammar WFST (SG-WFST) from the unigram ($\phi$) and bigram ($\mathbb{B}$) conditional densities over the subwords. Any valid path in SG-WFST gives a sequence of subwords and the joint probability associated with that sequence is given by the cost incurred in traversing along that path. Example SG-WFSTs for the given parameters $\phi$ and $\mathbb{B}$ are shown in Figs. ~\ref{fig_3_1} and ~\ref{sg_wfst} for Tamil and Kannada respectively.

\begin{figure*}[!ht]
\includegraphics[width=\textwidth,height=0.5\textheight]{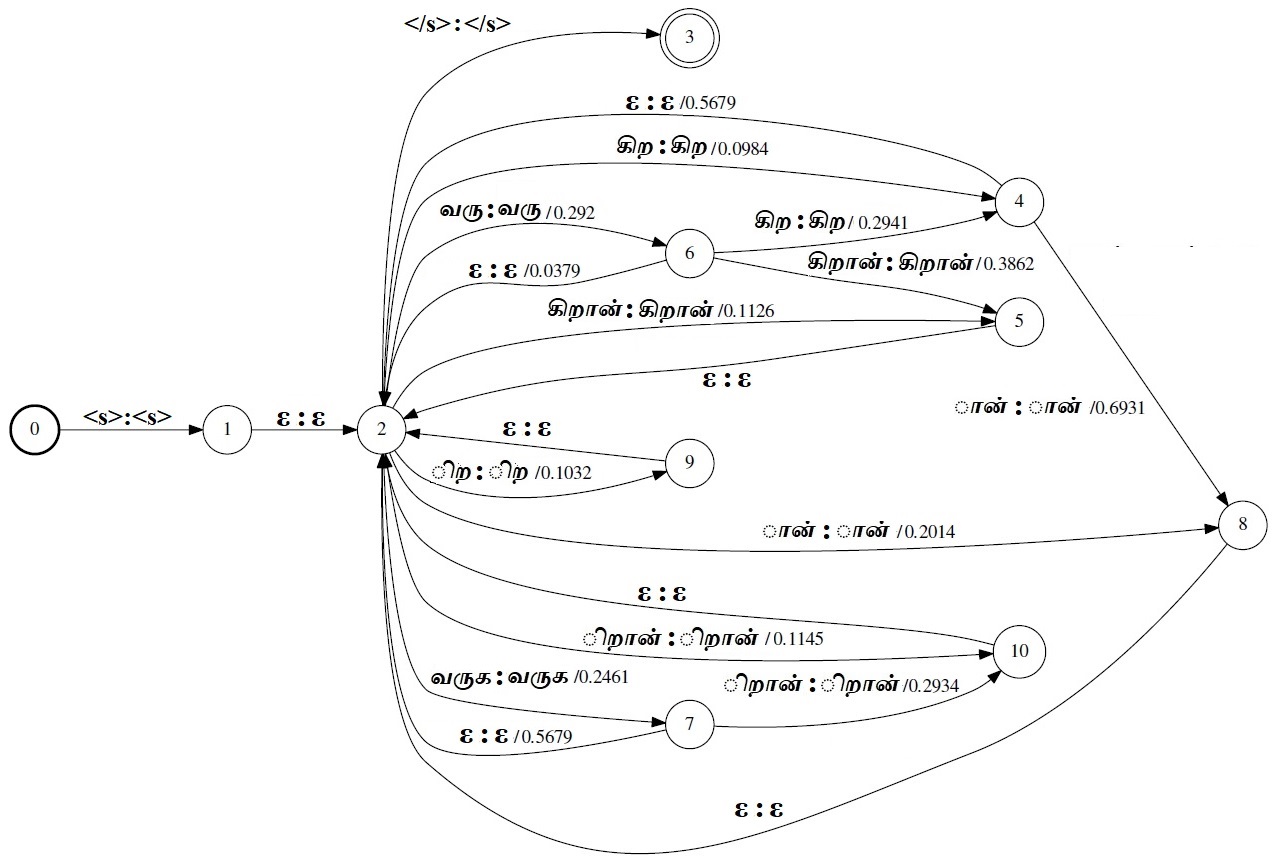}
\centering
\caption{A sample Tamil subword grammar weighted finite state transducer (SG-WFST) generated for the subword dictionary $\mathbb{D}$ shown in figure \ref{fig_3_1}a.}
	\label{fig_3_2}
\end{figure*}

\subsection*{\textbf{Construction of W-WFST :}}
\begin{figure}[!ht]
\includegraphics[width=\textwidth,height=0.24\textheight]{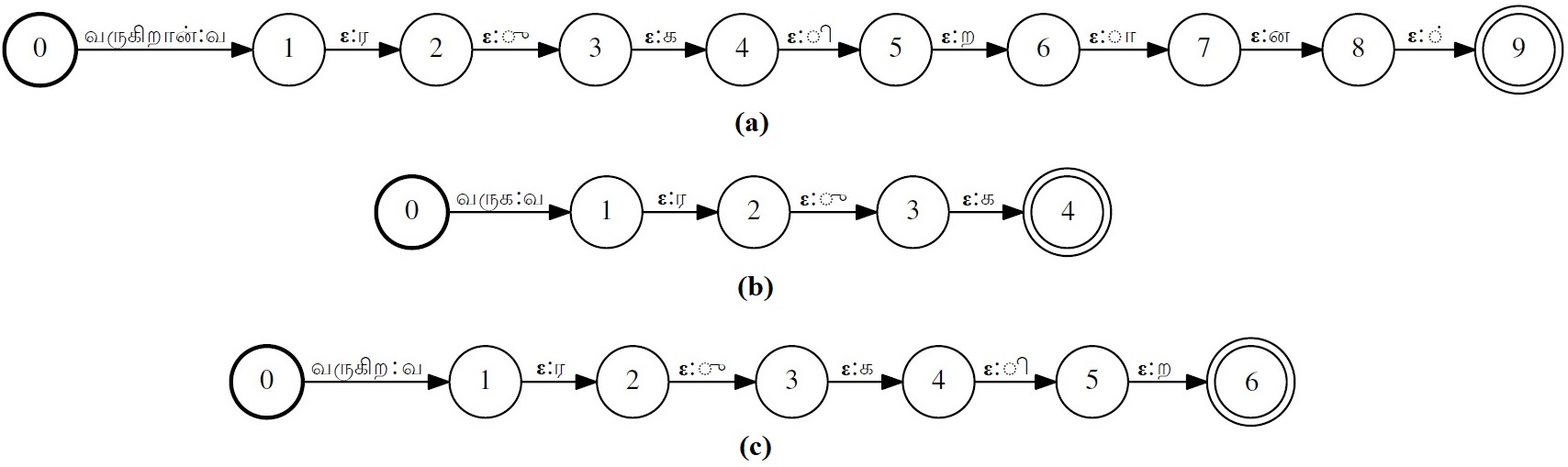}
\centering
\caption{Word weighted finite state transducer (W-WFST) generated for three Tamil words (a) \protect\scalerel*{\includegraphics{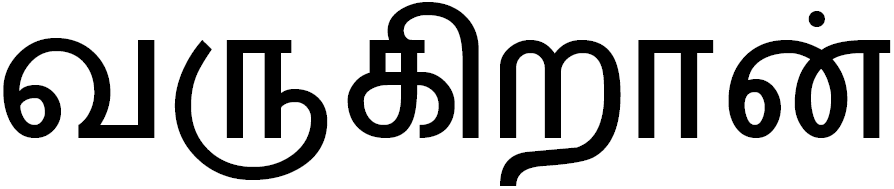}}{B} /varugiraan/ (b) \protect\scalerel*{\includegraphics{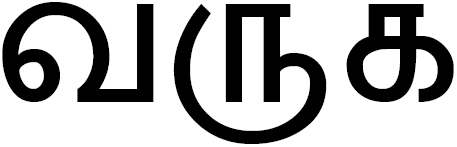}}{B} /varuga/ (c) \protect\scalerel*{\includegraphics{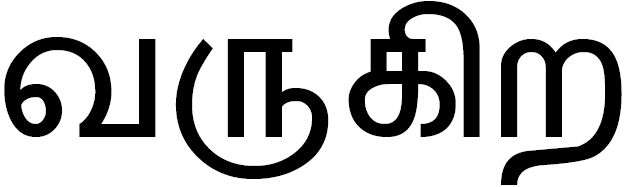}}{B} /varugira/}
	\label{fig_3_3}
\end{figure}

\begin{figure}[hbtp]
\centering
\includegraphics[scale= 0.40]{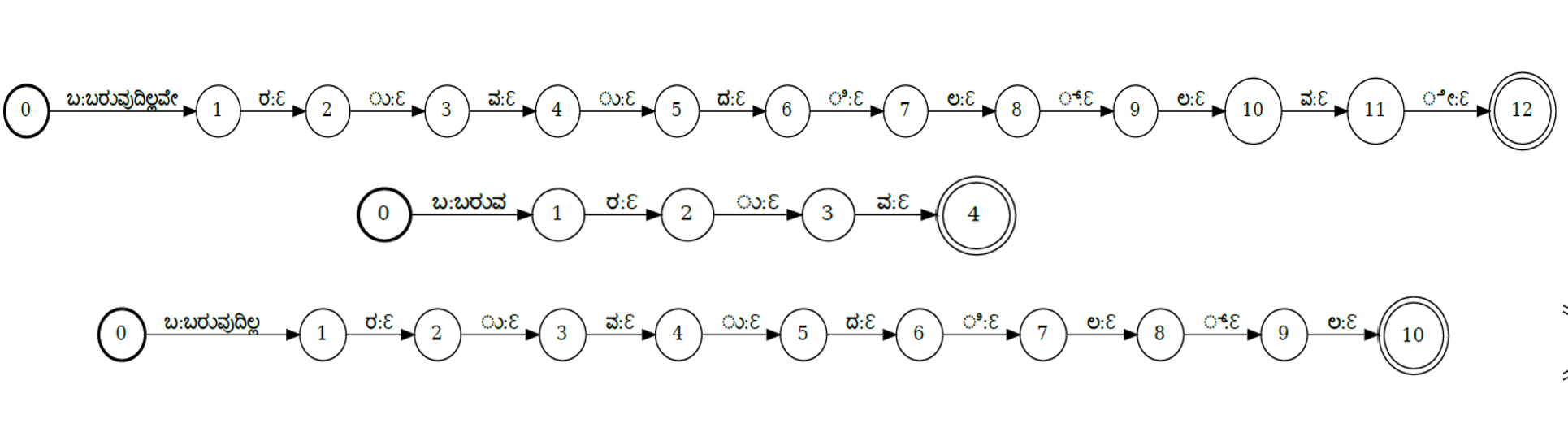}
\caption{Word weighted finite state transducer (W-WFST) generated for three Kannada words (a) \protect\scalerel*{\includegraphics{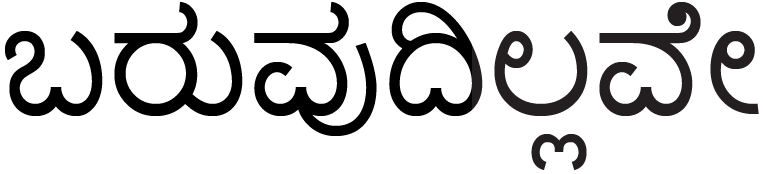}}{B} /baruvudillave/    (b) \protect\scalerel*{\includegraphics{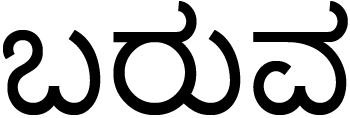}}{B}   /baruva/    and (c) \protect\scalerel*{\includegraphics{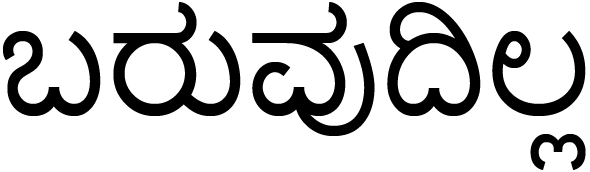}}{B}  /baruvudilla/    }
\label{w_wfst}
\end{figure}

For each word $w$ in $\mathbb{V}$, we construct a W-WFST which contains only one path having unique start and end states. The input label for the first transition of the path is the word $w$ and $\epsilon$ for the rest of the transitions, while the output labels are the individual characters of $w$. Figures \ref{fig_3_3} and ~\ref{w_wfst} show the W-WFSTs for some sample words for Tamil and Kannada respectively. 

\subsection*{\textbf{Segmentation by searching through O-WFST:}} To segment a word, we take its corresponding W-WFST and left-compose it with SD-WFST and SG-WFST. The resulting WFST is minimized, topology sorted and output-label projected to get O-WFST:

\begin{equation}
    \label{eqn_3_71}
    O = project_{output} \left ( topsort \left( min \left ( W \circ SD \circ SG \right) \right ) \right )
\end{equation}

\begin{figure*}[!h]
\includegraphics[width=\textwidth,height=0.42\textheight]{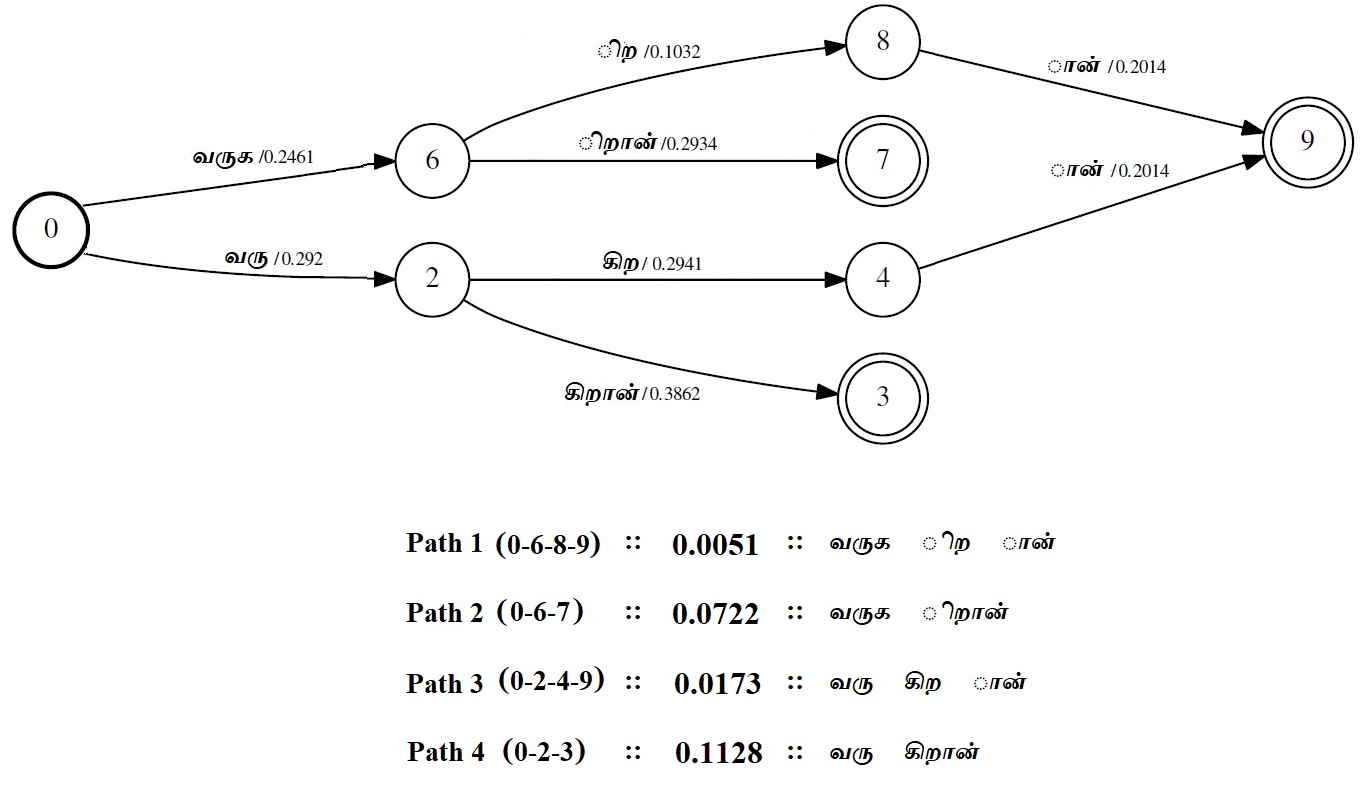}
\centering
\caption{O-WFST obtained by composing the W-WFST for a sample Tamil word in figure \ref{fig_3_3}a with SG-WFST in figure \ref{fig_3_2} (showing possible segmentation paths and their corresponding weights)}
	\label{fig_3_4}
\end{figure*}

\begin{figure*}[!h]
\includegraphics[width=\textwidth,height=0.42\textheight]{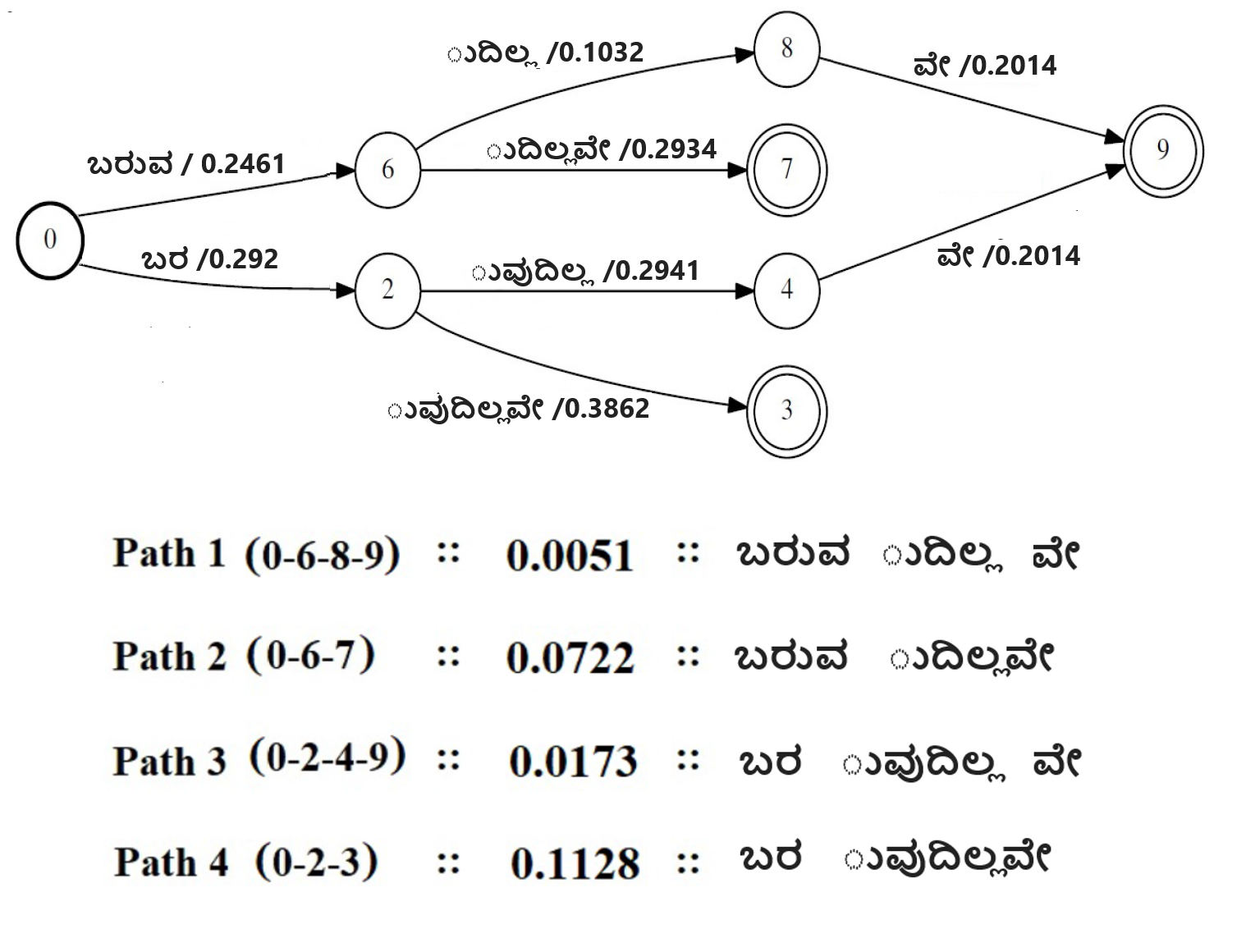}
\centering
\caption{O-WFST obtained by composing the W-WFST for a sample Kannada word in figure \ref{w_wfst}a with SG-WFST in figure \ref{fig_3_2} (showing possible segmentation paths and their corresponding weights)}
	\label{fig_owfst_kan}
\end{figure*}

The O-WFST thus obtained is shown respectively in figures \ref{fig_3_4} and \ref{fig_owfst_kan} for an example Tamil and Kannada word $w$, which have four different paths from the start to the end state, denoting four possible segmentation for $w$. We take the output label sequences across all these paths and create the list $\mathbb{W}_w$. The posterior value $\gamma_k(\overline{Z})$ corresponding to a segmentation sequence $\overline{Z}$ is the weight associated with that particular path that encodes the sequence $\overline{Z}$. The subword sequence corresponding to the path with the maximum weight is the maximum likelihood subword sequence $\overline{Z}_{w}^{*}$.

\begin{algorithm}[!ht]
\caption{Algorithm for ML-based word segmentation.}
\label{alg_3}
\SetAlgoLined
\vspace{5pt}
\KwIn{\hspace{0.2cm} (i) Subword dictionary $\mathbb{D}$ \\
      \hspace{1.65cm} (ii) Subword unigram counts $\mathbb{U}$ \\
      \hspace{1.65cm} (iii) Word vocabulary $\mathbb{V}$
}
\vspace{5pt}
\KwOut{
(i) List of segmented words $\left \{ \overline{Z}_{w}^{*} : \forall w \in \mathbb{V} \right \}$
}
\vspace{5pt}
 Normalize $\mathbb{U}$ to get the initial unigram density function $\phi_0$; \\
 
 Initialize $\mathbb{B}_0$ to be uniform conditional densities; \\
 
 Construct W-WFST for all the words in $\mathbb{V}$; \\
 
 Construct SD-WFST using the subwords in $\mathbb{D}$; \\
 
 Construct SG-WFST with weights taken from $\phi_0$ and $\mathbb{B}_0$; \\
 
 \vspace{5pt}
 
 \For{$iter \gets 1 \textnormal{ to } 15$}{
   \For{$w \gets 1 \textnormal{ to } |\mathbb{V}|$}{
     Obtain O-WFST for word $w$ : $O_w = project( topsort( min( W_w \circ SD \circ SG ) ) )$; \\
     
     Obtain all the segmentations and their weights, $ \{ \overline{Z} \}_w , \{ \gamma \}_w $ by searching $O_w$
     
     Obtain best segmentation $\overline{Z}_{w}^*$ corresponding to the maximum weight in $\{ \gamma \}_w$; \\
   }
   
   \vspace{5pt}
   
   Estimate $\phi_{iter}$ and $\mathbb{B}_{iter}$ using $ \{ \{ \overline{Z} \}_{w} \: : \forall w \in \mathbb{V} \}$ and $\{ \{ \gamma \}_{w} \: : \forall w \in \mathbb{V} \}$; \\
   
   Construct SG-WFST using subword dictionary $\mathbb{D}$ and the updated model parameters $\phi_{iter}$ and $\mathbb{B}_{iter}$; \\
 }
 
 \vspace{5pt}
 
 \textbf{return } $\{\overline{Z}^*\}_{w=1:|\mathbb{V}|}$ obtained at the last iteration;
\end{algorithm}

Once we segment all the words during an iteration $k$, we obtain the estimates $\phi$ and $\mathbb{B}$ for the next iteration $k+1$ and update the weights in SG-WFST with these new estimates and segment all the words again. These segmentation and SG-WFST update steps are performed for 15 iterations and we use the $\overline{Z}_{w}^{*}$ obtained after the last iteration to be the final ML-segmented subword sequence for the word $w$. The pseudocode to implement the ML-based word segmentation is given in Algorithm \ref{alg_3}.

\section{WFST implementation of Viterbi word segmentation}
\label{sec:Viterbi}

\begin{figure*}[!ht]
\includegraphics[width=0.53\textwidth,height=0.4\textheight]{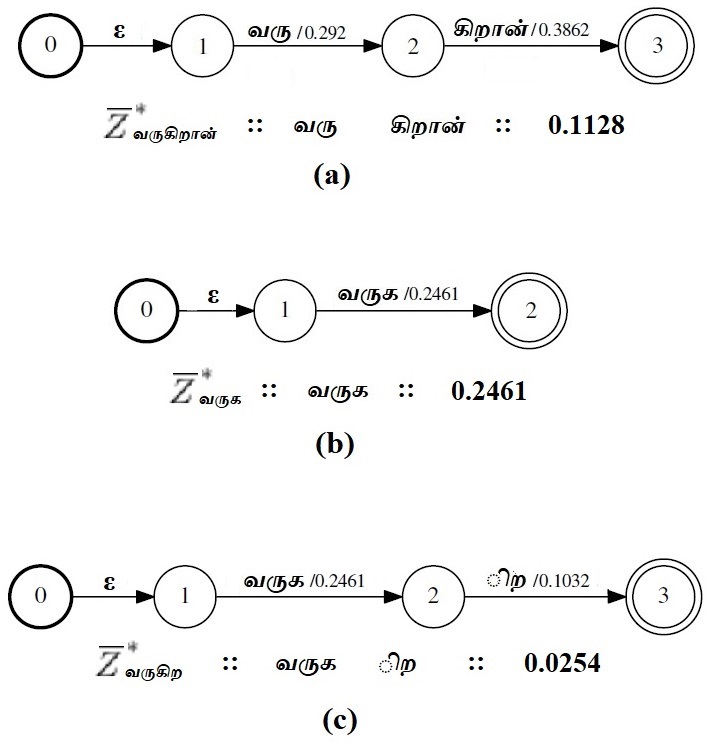}
\centering
\caption{Optimal segmentation paths (and their corresponding weights) obtained using Viterbi-based word segmentation for the W-WFSTs shown in figure \ref{fig_3_3} for some sample Tamil words.}
	\label{fig_3_5M}
\end{figure*}

The implementation of Viterbi word segmentation is almost similar to that of its ML counterpart. The only difference is that, while estimating the parameters using equations \ref{eqn_69} and \ref{eqn_70}, we consider only one segmentation path in O-WFST that has maximum weight instead of using all possible segmentation paths. Figure \ref{fig_3_5M} shows the O-WFST containing the best segmentation path obtained using Viterbi-based segmentation for the three W-WFSTs for 3 example Tamil words shown in figure \ref{fig_3_3} with the same subword dictionary and model parameters ($\phi$ and $\mathbb{B}$) shown in figures \ref{fig_3_1} and \ref{fig_3_2}. The procedural implementation of Viterbi word segmentation is given in Algorithm \ref{alg_4}.

\begin{algorithm}[!ht]
\caption{Algorithm for Viterbi-based word segmentation}
\label{alg_4}
\SetAlgoLined
\vspace{5pt}
\KwIn{\hspace{0.3cm} (i) Subword dictionary $\mathbb{D}$ \\
      \hspace{1.65cm} (ii) Subword unigram counts $\mathbb{U}$ \\
      \hspace{1.65cm} (iii) Word vocabulary $\mathbb{V}$
}
\vspace{5pt}
\KwOut{
(i) List of segmented words $\left \{ \overline{Z}_{w}^{*} : \forall w \in \mathbb{V} \right \}$
}
\vspace{5pt}
 Normalize $\mathbb{U}$ to get the initial unigram density function $\phi_0$; \\
 
 Initialize $\mathbb{B}_0$ to be uniform conditional densities; \\
 
 Construct W-WFST for all the words in $\mathbb{V}$; \\
 
 Construct SD-WFST using the subwords in $\mathbb{D}$; \\
 
 Construct SG-WFST with weights taken from $\phi_0$ and $\mathbb{B}_0$; \\
 
 \vspace{5pt}
 
 \For{$iter \gets 1 \textnormal{ to } 15$}{
   \For{$w \gets 1 \textnormal{ to } |\mathbb{V}|$}{
     Obtain O-WFST for word $w$ : $O_w = project( topsort( min( W_w \circ SD \circ SG ) ) )$; \\
     
     Obtain best segmentation $\overline{Z}_{w}^*$ corresponding to the maximum weighted path in $O_w$ WFST; \\
     
   }
   
   \vspace{5pt}
   
   Estimate $\phi_{iter}$ and $\mathbb{B}_{iter}$ by calculating the counts of subword occurrences in $\left \{ \overline{Z}_{w}^{*} : \forall w \in \mathbb{V} \right \}$; \\
   
   Construct SG-WFST using subword dictionary $\mathbb{D}$ and the updated model parameters $\phi_{iter}$ and $\mathbb{B}_{iter}$; \\
 }
 \vspace{5pt}
 
 \textbf{return } $\left \{ \overline{Z}_{w}^{*} : \forall w \in \mathbb{V} \right \}$ obtained at the last iteration;
 
\end{algorithm}

\section{Construction of subword based ASR}
\label{sec:subASR}
In this section, we explain the steps involved in building the subword-ASR.
\begin{itemize}
\item Segment the words in transcription text and in the LM text corpus into subwords using one of the combinations of the dictionary creation and segmentation techniques.

\item Add context markers ($+$) to the segmented subwords to differentiate between prefixes, suffixes, infixes or singleton words as explained in \cite{sub_context}.

\item Learn N-gram subword LM from the segmented text corpus.

\item Collect all the context-marked subwords and construct the lexicon WFST similar to that of the baseline word-based ASR system.

\item Compose the subword lexicon WFST and the subword grammar WFST and use it to train and test the ASR system.

\item During testing, post-process the output subword sequence by combining the subwords back to words by removing the context-markers \cite{sub_context}.
\end{itemize}

We have used $6^{th}$ order LM to learn the grammar of subword units in all our ASR experiments. The reason to have $6^{th}$ order subword LM compared to 3rd order word level LM is because each word is normally split into 2-4 subwords and so only by increasing the subword LM order to 6, we are able to match the performance and learn the grammar context of the 3rd order word-level LM. We have also varied the order of the subword-LM from 2 to 6 and compared their WER performances.

We have considered only the words in the training corpus to build lexicon for the baseline ASR for the purpose of comparison and to illustrate the benefits of subword-ASR.

\section{Experimental setup and results}
\label{sec:exp}

In this section, we compare the performances of subword-ASRs with the baseline ASR system in terms of OOV rate and WER and empirically justify the need for subword modeling to handle highly agglutinative languages like Tamil and Kannada. OOV rate is defined as the ratio of number of words in the test data which are not present in the training corpus to the total number of words in the test data.

We have tried different combinations of automatic subword dictionary methods (\textit{Morfessor}, BPE and extended-BPE) with ML and Viterbi-based segmentation methods and evaluated the WER performance of the subword-ASR for different orders of subword n-gram LMs with the dataset explained in section \ref{secDataBaseline}. In our experiments, the subword dictionary size is chosen to be 20000 for BPE method. For extended-BPE method, we choose $N_1=48$, $N_2=1000$, $N_3=4000$, $N_4=6000$, $N_5=4000$, $N_6=3000$ and $N_7=1952$ so that the total subword dictionary size is 20000. These values are chosen after various trial and error experiments such that we get the best possible recognition accuracy, while maintaining the size of the subword ASR model almost the same as that of the word-based ASR model. For morphological analyzer-based dictionary creation method, the size of the subword dictionary is automatically decided by the \textit{Morfessor} tool (in our case, the size comes out to be 22834). The baseline word-based ASR system has a total vocabulary size of 182771 words for Tamil and 201055 words for Kannada.

\subsection{Performance of ML segmentation technique}

Tables \ref{tab_MLTamil} and \ref{tab_MLKannada} compare the WER performance of the proposed subword dictionary creation method for Tamil and Kannada respectively, using ML-based segmentation with the baseline system for various orders of LM. We have also listed the OOV rates to further justify the effectiveness of the subword-ASR system.

We see that all the subword dictionary creation methods using ML segmentation perform better than the baseline word-based model, for both the languages. We obtain WERs of 15.03\% and 14.42\% for \textit{Morfessor} and BPE-based dictionary creation, respectively, for Tamil. The best WER of 14.10\% is obtained by the extended-BPE for $6^{th}$ order subword LM which is an absolute improvement of 10.6\% over the baseline model. For Kannada, we obtain WERs of 13.02\% and 12.83\% for \textit{Morfessor} and BPE based dictionary creation methods respectively, with ML-segmentation. Similar to Tamil, extended-BPE method gives the best WER of 12.31\%, which is an absolute WER improvement of 9.64\% over the baseline Kannada ASR system.

\begin{table}[ht]
\centering 
  \caption{Comparison of the word error rates and out of vocabulary rate of different subword dictionary creation techniques with maximum likelihood segmentation for Tamil.}
  \begin{tabular}{| c | c| c| c |c |c| }
  \hline
    \multirow{3}{*}{Method} & \multicolumn{4}{c|}{Word Error Rate (\%)} & \multirow{3}{*}{OOV} \\[2pt]
	\cline{2-5}
	& 3-gram & 4-gram & 5-gram & 6-gram &
	\\[2pt]
	& LM & LM & LM & LM &
		\\[4pt]
	\hline
	\hline
	Baseline (word-based) & 24.70 & -NA- & -NA- & -NA- & 10.73\\[4pt]
	\hline
	Morfessor & 18.84 & 18.29 & 15.47 & 15.03 & 2.41\\[4pt]
	\hline
	Byte pair encoding & 17.97 & 17.58 & 15.24 & 14.42 & 0.0  \\[4pt]
	\hline
	Extended-BPE & 17.43 & 16.72 & 14.96 & 14.10 &  0.0 \\[4pt]
	\hline
\end{tabular}%
\label{tab_MLTamil}
\end{table}

\begin{table}[ht]
\centering 
  \caption{Comparison of the word error rates and out of vocabulary rate of different subword dictionary creation techniques with maximum likelihood segmentation for Kannada.}
  \begin{tabular}{| c | c |c| c| c |c| }
  \hline
    \multirow{3}{*}{Method} & \multicolumn{4}{c|}{Word Error Rate (\%)} & \multirow{3}{*}{OOV} \\[2pt]
	\cline{2-5}
	& 3-gram & 4-gram & 5-gram & 6-gram &
	\\[2pt]
	& LM & LM & LM & LM &
	\\[4pt] 
	\hline
	\hline
	Baseline (word-based) & 21.95 & -NA- & -NA- & -NA- & 8.64 \\[4pt]
	\hline
	Morfessor & 16.02	& 15.30&	14.22	&13.02	&2.06\\[4pt]
	\hline
	Byte pair encoding & 15.94	&14.13&	13.24&	12.83	&0 \\[4pt]
	\hline
	Extended-BPE & 	15.71	&14.01&	12.98&	12.31&	0 \\[4pt]
	\hline
\end{tabular}%
\label{tab_MLKannada}
\end{table}   

Such huge improvements in WER of subword ASR are due to the reduction in the OOV rate. We see that BPE and extended-BPE methods have 0\% OOV rate. The reason for 0\% OOV rate in BPE-based methods (for both Tamil and Kannada) is that their subword dictionaries contain all the individual Tamil or Kannada characters as subwords; hence, every possible word can be segmented using these subword dictionaries.

\subsection{Performance of Viterbi segmentation technique}
We compare the WER performance of the subword dictionary creation methods using Viterbi segmentation for various LM orders with the baseline system for Tamil and Kannada in tables \ref{tab_ViterbiTamil} and \ref{tab_ViterbiKannada} respectively. For Tamil, the \textit{Morfessor} and BPE-based subword dictionary creation methods give WERs of 15.81\% and 15.13\%, respectively. The extended-BPE method performs the best with a WER of 14.97\% for $6^{th}$ order subword LM, which is an absolute WER improvement of 9.73\% over the baseline model. For Kannada, the \textit{Morfessor} and BPE-based subword dictionary creation methods give WERs of 13.45\% and 13.16\% respectively, whereas extended-BPE performs the best with a WER of 12.82\% obtained for $6^{th}$ order subword LM (an absolute WER improvement of 9.13\% over the baseline model).

The OOV rates for the Viterbi segmentation technique show a trend similar to that of ML-segmentation, but slightly poorer. We see that BPE and extended-BPE methods have zero OOV rates whereas the baseline models have an OOV rate of 10.73\% and 8.64\% for Tamil and Kannada respectively.

\begin{table}[ht]
\centering 
  \caption{Comparison of the word error rates and out of vocabulary rates of the subword dictionary creation with Viterbi segmentation for Tamil.}
  \begin{tabular}{| c | c| c| c| c | c| }
  \hline
     \multirow{3}{*}{Method} & \multicolumn{4}{c|}{Word Error Rate (\%)} & \multirow{3}{*}{OOV} \\[2pt]
	\cline{2-5}
	& 3-gram & 4-gram & 5-gram & 6-gram & 
	\\[2pt]
	& LM & LM & LM & LM &
	\\[4pt]
	\hline
	\hline
	Baseline (word-based) & 24.70 & -NA- & -NA- & -NA- & 10.73\\[4pt]
	\hline
	Morfessor & 22.53 & 20.08 & 16.14 & 15.81 & 3.76 \\[4pt]
	\hline
	Byte pair encoding & 18.74 & 17.88 & 15.62 & 15.13 &  0.0\\[4pt]
	\hline
	Extended-BPE & 18.46 & 17.51 & 15.28 & 14.97 & 0.0\\[4pt]
	\hline
\end{tabular}%
\label{tab_ViterbiTamil}
\end{table}

\begin{table}[ht]
\centering 
  \caption{Comparison of the word error rates and out of vocabulary rates of the subword dictionary creation with Viterbi segmentation for Kannada.}
  \begin{tabular}{| c | c |c| c| c |c| }
  \hline
    \multirow{3}{*}{Method} & \multicolumn{4}{c|}{Word Error Rate (\%)} & \multirow{3}{*}{OOV} \\[2pt]
	\cline{2-5}
	& 3-gram & 4-gram & 5-gram & 6-gram & 
	\\[2pt]
	& LM & LM & LM & LM &
	\\[4pt] 
	\hline
	\hline
	Baseline (word-based) & 21.95 & -NA- & -NA- & -NA- & 8.64 \\[4pt]
	\hline
	Morfessor & 16.82 &	15.98 &	14.72 &	13. 45 &	2.92 \\[4pt]
	\hline
	Byte pair encoding & 15.86 &	14.95 &	13.62 &	13.16 &	0 \\[4pt]
	\hline
	Extended-BPE & 	15.32 &	14.21 &	13.14 &	12.82 &	0 \\[4pt]
	\hline
\end{tabular}%
\label{tab_ViterbiKannada}
\end{table}   

Comparing the segmentation techniques, we observe that ML segmentation fares better than Viterbi segmentation method for both languages, which is due to the fact that the former soft-weighs all the possible segmentation paths and estimates the subword model parameters while the latter considers only the best segmentation path.

\section{Conclusion}
\label{sec:conc}
Thus, we have presented ASR systems based on novel subword modeling techniques to handle the infinite vocabulary problem of highly agglutinative languages like Tamil and Kannada. We have explored BPE, extended BPE and \textit{Morfessor} tool to construct subword dictionaries and have used statistical approaches such as ML and Viterbi to segment words into subwords efficiently using WFST framework. The experiments demonstrated the success of the proposed approach in terms of reduction in WER and OOV rates. We  also notice that the ML method performs slightly better than the Viterbi method for both the languages.

\end{document}